\documentclass[conference]{IEEEtran}
\IEEEoverridecommandlockouts
\usepackage{cite}
\usepackage{amsmath,amssymb,amsfonts}
\usepackage{algorithm}
\usepackage{algorithmic}
\usepackage{graphicx}
\usepackage{textcomp}
\usepackage{xcolor}
\usepackage{balance}
\usepackage{siunitx}
\usepackage{caption}
\usepackage{subcaption}
\usepackage{multirow}
\usepackage{makecell}
\usepackage{optidef}
\usepackage[a4paper,
            left=0.75in,
            right=0.75in,
            top=0.7in,
            bottom=1in]{geometry}

\DeclareMathOperator*{\argmax}{arg\,max}

\def\BibTeX{{\rm B\kern-.05em{\sc i\kern-.025em b}\kern-.08em
    T\kern-.1667em\lower.7ex\hbox{E}\kern-.125emX}}
\begin{document}

\title{Beam Selection in ISAC using Contextual Bandit with Multi-modal Transformer and Transfer Learning \vspace{-1ex}}

\author{
\normalsize
Mohammad Farzanullah\IEEEauthorrefmark{1}, Han Zhang\IEEEauthorrefmark{1}, Akram Bin Sediq\IEEEauthorrefmark{2}, Ali Afana\IEEEauthorrefmark{2} and Melike Erol-Kantarci\IEEEauthorrefmark{1}\\
\vspace{0ex}
\IEEEauthorblockA{\IEEEauthorrefmark{1}
    School of Electrical Engineering and Computer Science, University of Ottawa, Ottawa, ON, Canada} 
\IEEEauthorblockA{\IEEEauthorrefmark{2}
    Ericsson Inc., Ottawa, ON, Canada}
Emails: \{mfarz086, hzhan363, melike.erolkantarci\}@uottawa.ca
\\ \{akram.bin.sediq, ali.afana\}@ericsson.com
\vspace{-3ex}
} 

\maketitle

\begin{abstract}
Sixth generation (6G) wireless technology is anticipated to introduce Integrated Sensing and Communication (ISAC) as a transformative paradigm. ISAC unifies wireless communication and RADAR or other forms of sensing to optimize spectral and hardware resources. This paper presents a pioneering framework that leverages ISAC sensing data to enhance beam selection processes in complex indoor environments. By integrating multi-modal transformer models with a multi-agent contextual bandit algorithm, our approach utilizes ISAC sensing data to improve communication performance and achieves high spectral efficiency (SE). Specifically, the multi-modal transformer can capture inter-modal relationships, enhancing model generalization across diverse scenarios. Experimental evaluations on the DeepSense 6G dataset demonstrate that our model outperforms traditional deep reinforcement learning (DRL) methods, achieving superior beam prediction accuracy and adaptability. In the single-user scenario, we achieve an average SE regret improvement of 49.6\% as compared to DRL. Furthermore, we employ transfer reinforcement learning to reduce training time and improve model performance in multi-user environments. In the multi-user scenario, this approach enhances the average SE regret, which is a measure to demonstrate how far the learned policy is from the optimal SE policy, by 19.7\% compared to training from scratch, even when the latter is trained 100 times longer. 
\end{abstract}


\vspace{-1ex}
\section{Introduction}

The sixth generation (6G) of wireless technology aims to support numerous new intelligent applications
and meet the requirements of extremely high data rates. Integrated Sensing and Communication (ISAC) has been identified as one of the key areas in 6G communications
\cite{lu2024integrated}. 
ISAC is an emerging paradigm that seeks to unify the traditionally separate functions of wireless communication and sensing into a single framework. This integration aims to optimize the use of spectral and hardware resources and enable devices to simultaneously transmit data and sense the environment. 
On one hand, by leveraging shared infrastructure and spectrum, ISAC systems can enhance situational awareness, improve communication reliability, and support new applications like autonomous driving and smart cities. On the other hand, millimeter wave (mmWave) communication requires precise beam selection to achieve high data rates.
However, selecting the precise beam requires high training overhead, which makes it less suitable for mobile environments.
A key point is that mmWave communication's directional nature means beam management depends on the position of the antenna, user equipment (UE), and the surrounding environment. Thus, sensing user location and environment geometry can help guide beam management and reduce training overhead \cite{jiang2022lidar}. Some recent works have demonstrated the effectiveness of LiDAR \cite{jiang2022lidar}, RADAR \cite{demirhan2022radar}, and camera \cite{charan2022vision} for beam prediction task. However, the installation of these sensors can impose an extra cost, and they suffer from instability. For instance, cameras are less reliable in adverse weather conditions,
and LiDAR sensors can be expensive \cite{Erol2412:Generative}. While future BSs are expected to have ISAC sensing capability, its effectiveness for beam prediction is yet to be evaluated. 

For beamforming in ISAC systems, several studies have focused on beamforming for jointly optimizing sensing and communication capabilities.  \cite{choi2024joint} designs a framework for beamforming optimization to maximize spectral efficiency while maintaining sensing performance in a multi-user and multi-target ISAC system. 
Meanwhile, \cite{hua2023optimal} studies transmit beamforming in a system integrating downlink communication and sensing, optimizing sensing performance while ensuring communication SINR requirements.
However, these works do not consider how the presence of ISAC sensing information can further improve communication performance. 
As compared to the existing studies, our work examines how the ISAC sensing capability can help with beam selection and improve the spectral efficiency.

Inspired by these ideas, we design a novel framework by integrating transformers with contextual bandits and perform beam selection based on ISAC data.
Transformer models \cite{vaswani2017attention} constitute a neural network architecture that excels in handling sequential data through the use of self-attention mechanisms, enabling them to capture long-range dependencies.
Furthermore, a multi-modal transformer (MMT) encoder is an extension of the transformer architecture tailored for processing multi-modal data \cite{xu2023multimodal}. While traditional transformers employ self-attention mechanisms that enable each token to focus on other tokens within the same input sequence, a multi-modal configuration incorporates cross-modal attention mechanisms. This facilitates the model's ability to learn relationships between modalities.
A few works utilize multi-modal learning for beam prediction \cite{charan2022vision}; however, they do not take advantage of the MMT architecture.

Contextual bandits \cite{lattimore2020bandit} 
is a type of reinforcement learning (RL) algorithm that extends the multi-armed bandit problem by incorporating context or additional information to make more informed decisions. In the traditional multi-armed bandit problem, an agent must choose between different options to maximize rewards, learning over time which options yield the best results. Contextual bandits enhance this framework by considering the context in which decisions are made
allowing the algorithm to tailor its choices based on the specific situation. Unlike other RL approaches, 
contextual bandits focus on optimizing immediate rewards without considering future states. Recently, transformer models have been utilized in RL tasks \cite{li2023survey}. 
Their ability to model complex dependencies enhances their effectiveness in RL, improving policy learning and decision-making.
In a recent study \cite{Ghassemi}, a two-stage process employing MMT and single agent RL was utilized for beam selection. However, this approach differs from ours as it does not integrate MMT within the RL agent nor utilize ISAC data for prediction.

In this work, we utilize MMT in multi-agent contextual bandit for beam selection for multiple mobile UEs.
The main contributions can be summarized as follows:
\begin{itemize}
    \item To the best of our knowledge, we are the first to use ISAC sensing data to enhance beam 
    selection, demonstrating its potential to significantly enhance spectral efficiency in dynamic 6G environments.
    Compared with other sensing modalities \cite{jiang2022lidar, demirhan2022radar, charan2022vision}, using ISAC sensing avoids installation of sensors at the base station, and reduces extra cost.
    \item We propose a novel framework that integrates an MMT encoder with a multi-agent RL algorithm. The MMT encoder enables the model to learn inter-modal relationships between ISAC sensing data and user location data, improving generalization across diverse scenarios. By combining these two structures, this framework is able to perform cooperative beam selection, effectively addressing the challenge of mutual interference in dynamic multi-user environments.
    \item We demonstrate the effectiveness of transfer learning (TL) to adapt RL model trained in single-user scenarios to multi-user environments, significantly reducing training time while improving beam selection accuracy.
\end{itemize}

We evaluate our model on the DeepSense 6G dataset. We show that our model can learn to select beams that result in a high sum spectral efficiency for the users in the environment. Furthermore, the model generalizes well across diverse scenarios, adapting to varying environmental conditions, which highlights its robustness and flexibility for dynamic 6G networks.

The rest of the paper is organized as follows: Section \ref{Section:SysModel} highlights the system model and problem formulation. Next, the multi-agent contextual bandits algorithm is discussed in Section \ref{Section:Contextual}. This is followed by the simulation setup in Section \ref{Section:SimulationSetup}, and results in Section \ref{Section:Results}. Finally, we conclude the paper in Section \ref{Section:Conclusion}.

\section{System Model and Problem Formulation} \label{Section:SysModel}

In this work, we consider an indoor setting and the downlink communication scenario, with one Access Point (AP) equipped with a millimeter-wave (mmWave) phased array and multiple mobile UE devices. The AP is required to establish connectivity with multiple users. Let $K$ denote the number of users, and $\mathcal{K}$ denote the set of all users. Furthermore, consider there are $N$ antennas in the phased array, and the UEs have a single antenna. In this scenario, efficient beam selection is crucial to maintain high-quality communication links between the AP and the UEs. To achieve this, we employ a codebook-based beam selection strategy. 
Let $\{\textbf{f}_1, \textbf{f}_2, ....,  \textbf{f}_M\} \in \mathcal{F}$ denote the set of vectors in beamforming codebook, where each $\textbf{f}_i \in \mathbb{C}^{N\times1}$ (for $i = 1,2,...M$) is a complex vector of dimension $N \times 1$, and $M$ denotes the number of beamforming vectors present in the codebook. For each user $k \in \mathcal{K}$ being served by the $m_{th}$ beamforming vector $\textbf{f}_{m,k}$, the received signal at the user $k$ can be expressed as:
\begin{align}
    y_k = \sqrt{\frac{P_T}{K}} \textbf{h}_k^T \textbf{f}_{m,k}  x_k  + \sum_{\substack{k' \in \mathcal{K} \\ k' \neq k}} \sqrt{\frac{P_T}{K}} \textbf{h}_k^T \textbf{f}_{m,k'} x_{k'}  + z_k
\end{align}
where $P_T$ is the total transmit power, $x_k$ and $x_{k'}$ is the signal intended for user $k$ and $k'$ respectively, with $\mathbb{E}[|x_k|^2] = \mathbb{E}[|x_{k'}|^2] = 1$. Furthermore, $\textbf{h}_k^T \in \mathbb{C}^N$ is the transpose of the complex channel vector between AP and user $k$, and $z_k$ is the received additional white Gaussian noise (AWGN) at the UE $k$. The received signal at user $k$ will be impaired by the choice of beamforming vectors for other users $k'$.



Thus, the spectral efficiency of each user $k$ can be expressed using the Shannon capacity formula as \cite{sung2020multi}:
\begin{align}
    R_k = \text{log}_2 \left(1+ \frac{\frac{P_T}{K} ||\textbf{h}_k^T \textbf{f}_{m,k}||^2}{\sum_{\substack{k' \in \mathcal{K} \\ k' \neq k}} \frac{P_T}{K} ||\textbf{h}_k^T \textbf{f}_{m,k'}||^2 + \sigma^2} \right),
\end{align}
where $\sigma^2$ is the variance of the AWGN noise, and $||.||$ signifies the L2 norm of a vector.

\subsection{Problem Formulation}

The objective of the work is to determine the beams that would collaboratively maximize the sum spectral efficiency of all users moving around in the environment. For each user $k$ being served with a given beamforming vector $\textbf{f}_{m,k}$, we need to find the vector $\textbf{f}_{m,k}^*$ that would result in the highest sum spectral efficiency. The objective can be mathematically expressed as:
\begin{align}
      \textbf{f}_{m,1}^*, \textbf{f}_{m,2}^*, ....,  \textbf{f}_{m,K}^* = \argmax_{\{\textbf{f}_{m,k}\}_{k \in \mathcal{K}}} \quad & \sum_{k=1}^{K} R_k 
\end{align}
This represents a cooperative challenge, requiring the AP to assign beamforming vectors to all users in a manner that maximizes overall performance. The complexity of this scenario arises from the fact that the selection of a beamforming vector for one user may negatively impact the performance of other users. To address this issue, we propose a solution based on multi-agent contextual bandits, wherein the agents iteratively engage with the environment to learn an optimal policy, based on the multi-modal ISAC sensing and UE location data.

\section{Contextual Bandits based Beam Selection} \label{Section:Contextual}
\begin{figure*} [t]
    \centering
    \includegraphics[width=0.9\linewidth, trim=20pt 5pt 20pt 5pt, clip]{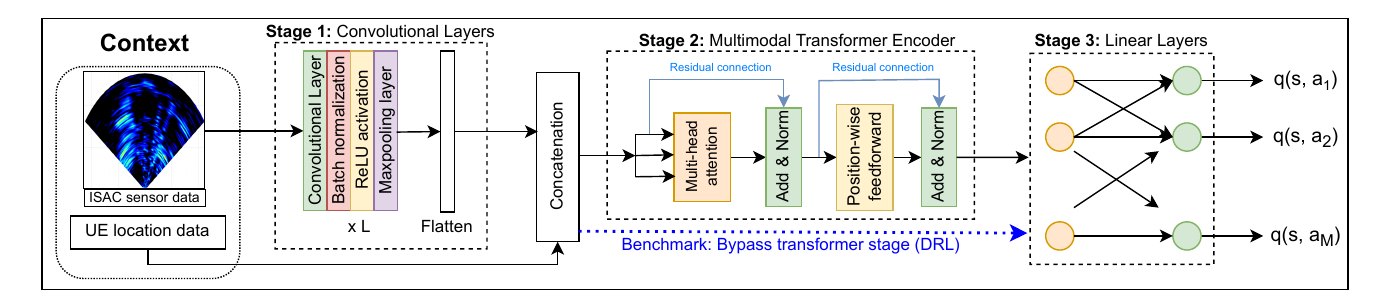}
    \caption{A contextual bandit agent. The context to the agent are the ISAC image and the UE location data, which are forwarded through three stages: Convolutional layers, MMT encoder, and linear layers. The output is the q-value for each action.}
    \label{fig:methodology}
\end{figure*}
In this section, we outline our proposed methodology. We model the problem as a multi-agent contextual bandits problem. Each UE acts as a separate agent, with the aim of selecting the beam that maximizes the sum spectral efficiency. 
To handle complex multi-modal data, we propose a novel transformer-based RL framework that integrates convolutional layers and a multi-modal transformer encoder into the contextual bandit framework. This integration leverages the power of the multi-modal transformer to learn relationships between different modalities. Additionally, the parameterized contextual bandit is employed to manage high-dimensional and complex state spaces for decision-making.

\subsection{Contextual Decision-Making Framework} \label{Section:ContextualDecisionFramework}
In this work, we use contextual bandits for decision-making since the state space at each time step is independent, and the selection of an action at time $t$ does not impact the future state $s_{t+1}$. During the training process, the agent needs to learn to select the best possible action $a_t$ given the context $s_t$ at each time step $t$.
The context, action space, and reward design are defined below:
\subsubsection{Context} For each UE $k$, the context $s_t$ at time step $t$ consists of the ISAC sensing data and the UE location data.
To the best of our knowledge, this is the first time ISAC sensing data is being used to enhance beam selection accuracy.
\subsubsection{Action space} 
The action space corresponds to the $M$ beamforming vectors. Let $\mathcal{M}$ denote the set of all actions. 
\subsubsection{Reward Design} 
To promote cooperative learning, we use a common reward for all agents.
The optimization goal of beam selection is to maximize the sum spectral efficiency. Therefore, the reward can be defined as the sum of spectral efficiency of all users, i.e., $\sum_{k=1}^{K} R_k$.
However, spectral efficiency is strongly dependent on distance.
This reward design causes agents to favor proximity-based actions, which leads to suboptimal decision-making in other scenarios, and results in unstable convergence during the training.
In this case, we propose a novel method for the reward design and introduce a distance-based factor in the reward formulation:
\begin{align} \label{Eq:reward}
    r_t = \sum_{k=1}^{K} R_k (d_k)^p
\end{align}
where $d_k$ is the distance of the UE $k$ from the AP at time $t$. $p$ is a hyperparameter that needs to be tuned empirically. With this design, the RL is able to achive more stable convergence.

\subsection{Multi-modal transformer-enhanced Contextual bandits}

Fig. \ref{fig:methodology} shows the architecture of our proposed multi-modal transformer-enhanced contextual bandit agent. The whole architecture can be divided into 3 parts. The context or state space is input to the agent. The ISAC data/image is forwarded through convolutional layers and then flattened. The UE location data are concatenated with the flattened output of convolutional stage, and then forwarded to the MMT encoder. 
The MMT encoder facilitates the model's ability to learn the relationships between the ISAC and UE location data.
Next, the output from the transformer encoder model is passed through linear neural layers to output the predicted q-value or reward for each action. 
In the following, we define the components of the model architecture.

\subsubsection{Convolutional layers}
The data received by the ISAC is processed as an image and passed through a sequence of $L$ convolutional layers. Each of these layers comprises a convolutional operation with $l_i$ filters, followed by batch normalization, a Rectified Linear Unit (ReLU) activation function, and a max-pooling layer. The purpose of convolutional layers is to automatically extract and learn hierarchical features from the input data, capturing spatial hierarchies and patterns that are crucial for effective image analysis. 
\subsubsection{Multi-modal Transformer Encoder}
A MMT encoder is employed to capture relationships in the multi-modal input data.
The output from stage 1, and the UE location data are concatenated and passed to the MMT encoder.
Multihead attention can perform multiple attention operations in parallel, enabling the model to simultaneously focus on various parts of the input sequence. The position-wise feedforward layer allows for localized non-linear transformations at each position within the input sequence, effectively capturing patterns.

\subsubsection{Linear Layers} 
The MMT encoder's output is processed through linear neural network layers, transforming encoded features into the final output. The output layer has a dimension $M$, corresponding to the number of beamforming vectors.

\subsection{Learning algorithm}
We focus on setting with multiple time steps $t$, where the state space $s_t$ at each time step $t$ is independent,
so each agent $k$ needs to learn the action $a_t^{(k)} \in \mathcal{M}$ to take given a context.

We adopt contextual bandits with experience replay to train multiple agents. 
We use $q_{\pi}(s_t, a_t) = \mathbb{E}[r_t]$ to define the reward or q-value for an agent when it selects an action $a_t \in \mathcal{M}$ following policy $\pi$, given the context $s_t$. The policy is defined as a mapping from a state to an action, represented by the conditional probability distribution $\pi (a|s)$. 
If the q-values of all actions can be accurately estimated, the optimal policy would be to simply select the action that results in the highest reward or q-value, i.e., $\argmax_{a_t \in \mathcal{M}} q(s_t, a_t)$. In our case, a parameterized network is used to approximate the q-values, as shown in Fig. \ref{fig:methodology}.

Each agent $k$ takes as input the context $s_t^{(k)}$, and outputs the q-value for each action. At each time step, the agents explore the state-action space with the $\epsilon$-greedy policy. 
The interactions of the agent are stored in a replay memory, and at fixed intervals $J$, a batch of interactions $\mathcal{D}$ are sampled to train the model. 
Replay memory enables to disrupt the correlation between consecutive time steps, stabilizing the learning process. The algorithm is presented in Algorithm 1. 
\begin{algorithm}
\caption{Beam selection with Contextual bandits}
\begin{algorithmic}[1]
\STATE Start environment simulator, generating UE and links
\FOR{each step $t$}
    \STATE Update UE locations and ISAC sensing data
    \FOR{each UE agent $k$}
        \STATE Observe $s_t^{(k)}$
        \STATE Choose action $a_t^{(k)} \in \mathcal{M}$ by $\epsilon$-greedy policy
    \ENDFOR
    \STATE All agents take actions and receive reward $r_{t}$
    \FOR{each UE agent $k$}
        \STATE Store $(s_t^{(k)}, a_t^{(k)}, r_{t})$ in the replay memory $\mathcal{D}_k$
    \ENDFOR
    \IF{$t \bmod J == 0$}
    \FOR{each UE agent $k$}
        \STATE Uniformly sample mini-batches from $\mathcal{D}_k$
        \STATE Optimize mean squared error (MSE) between q-value and $r_t$ using batch gradient descent
    \ENDFOR
    \ENDIF
\ENDFOR
\end{algorithmic}
\end{algorithm}

During the testing stage, at time step $t$, each trained agent $k$ observes the context $s_t^{(k)}$ and takes an action $a_t^{(k)}$.

\subsection{Transfer reinforcement learning}

In this study, we note that the complex architecture and large size of transformers necessitate significant amounts of training data and extended training periods. In this case, TL is an effective solution to accelerate model training and improve model accuracy.
In RL, TL uses knowledge from source tasks to enhance learning efficiency and performance in a target task, reducing the data and time needed to train models on new tasks by leveraging prior skills and experiences \cite{zhu2023transfer}.
It focuses on adapting learned policies to new environments, ensuring models generalize well across different tasks and scenarios. We utilize the model initially trained on the single-agent scenario and apply it to the multiple agent scenario. Subsequently, we continue training the multiple agents for an additional epoch. TL was implemented to show the generalization capabilities of the transformer-based model, and to reduce the training time of multiple agent case.

\section{Simulation Setup} \label{Section:SimulationSetup}

\subsection{Dataset}
We use the DeepSense 6G dataset and the Scenarios 42 and 44 within  \cite{DeepSense} focusing on an indoor environment. In Scenario 42, two users move across the room, whereas in Scenario 44, one user is considered. The dataset consists of different modality sensors placed at the AP, including camera, LiDAR, and RADAR. In our work, we use the ISAC sensing data and the location data for the beam selection of users. 

The ISAC data was collected by a phased array consisting of a 16-element uniform rectangular array. The correlation between the received waveform at the phased array and the transmitted waveform is calculated, which provides a 2D image or representation of the environment. 
The dataset does not include the locations of the UEs. To determine the user locations, we align the camera images with LiDAR data to accurately pinpoint their positions. Initially, we employ the pre-trained YOLOv5 model \cite{jocher2022ultralytics} to get the bounding boxes of users moving within the environment. These bounding boxes are used to calculate the azimuth angle $\phi$ and radial angle $\alpha$, which are applied to the LiDAR image to derive the users' exact coordinates. 
In a real-world scenario, if the exact locations of users are available (e.g., through GPS or other localization technologies), they can be used directly. 
As outlined in Section \ref{Section:ContextualDecisionFramework}, the ISAC data and the UE location serve as the context or input for the agent.

\subsection{Implementation Details}
For each scenario, we train our agents for the first 80\% of the data and reserve the last 20\% of the data for testing. Each scenario consists of 2100 time steps, with 100ms between each step. The AP has a uniform rectangular array with 2 vertical and 8 horizontal antennas. The carrier frequency of 60 MHz is used. Out of the predefined 64 beam codebook, we use the beams 12 to 49, that roughly approximate +- 30 degrees in the azimuth angle. 

We call our model Transformer Reinforcement Learning (TRL) for short, and compare our model with the following benchmarks: 1) We train the agents without using the MMT. In this case, the flattened output of convolutional layers and the position data are concatenated and directly forwarded to the stage 3. We call this model Deep Reinforcement Learning (DRL), shown by the blue dotted line in Fig. \ref{fig:methodology}.
2) Exhaustive Search algorithm: For each time step, all the action combinations for all agents are tested, and the actions that result in the highest sum spectral efficiency are selected. 
Specifically, for each time step, the algorithm iterates through all beamforming vectors in the codebook and calculates the sum spectral efficiency for all users. 
This represents the upper limit of the performance, serving as a benchmark for evaluating the effectiveness of our algorithm.
3) Random allocation: Each agent selects an action randomly.


An epoch is defined as the contextual bandit iterating over the training data once.
In the stage of the convolutional layer, we established $L=8$, with [8,8,8,8,8,4,4,1] filters in each layer. In the second stage, a single transformer encoder layer was utilized, incorporating a multi-head attention mechanism with 5 heads. The feedforward network model was configured with a dimensionality of 2048, and a dropout rate of 0.1 was applied. The third stage involved the implementation of two hidden layers, each containing 256 neurons, with a dropout rate of 0.4.
Furthermore, we reduced the value of $\epsilon$ from 1 to 0.02 for the first 80\% of the training stage, after which it was fixed at 0.02 for the rest of the training. MSE was used as the loss function, and Adam optimizer was used with a learning rate of 0.005. The value of $p$ in Eq. (\ref{Eq:reward}) is set as 0.4.


\section{Results and Discussion} \label{Section:Results}

In this section, we present the experimental results. 
We will first present the results for Scenario 44, which was a single UE case. Next, we will present results for Scenario 42, where two UEs are moving around in the environment. 

Apart from the spectral efficiency, we also use modified regret to show our model's proximity to the optimal policy. In the contextual bandits problem, regret is defined as the difference between the expected reward of the optimal policy and the expected reward of the policy chosen by the algorithm over a series of rounds, i.e., $X(T) = \sum_{t=1}^{T} (\mathbb{E}[r_t (a_t^*)] - \mathbb{E}[r_t (a_t)] )$, where $a_t^*$ is the optimal action that maximizes the expected reward, and $a_t$ is the action taken by the agent \cite{lattimore2020bandit}.

In our case, we define average spectral efficiency (SE)-regret, which is the average of the difference between the sum spectral efficiency obtained from a model, and the sum spectral efficiency obtained through exhaustive search, over a series of rounds. This is mathematically represented as:
\begin{align}
    X_{SE}(T) = \frac{1}{T} \sum_{t=1}^{T} \sum_{k=1}^{K} (R_k(a_t^{(k)*}) - R_k(a_t^{(k)}))
\end{align}
where $a_t^{(k)*}$ is the action selected by agent $k$ in the exhaustive search algorithm, representing the optimal policy. Meanwhile, $a_t^{(k)}$ is the action selected by the agent $k$ for the given model during the testing stage. The average SE regret measures how far our model is in terms of sum spectral efficiency from the policy of the optimal sum spectral efficiency.

\subsection{Scenario 44 (Single UE) results}
\vspace{0ex}
\begin{table} [h]
   \caption{Average SE regret, spectral efficiency, and optimal action selection comparison for Scenario 44 test data.
   }
   \centering \resizebox{0.95\columnwidth}{!}{
   \begin{tabular}{| c| c| c| c|}				
       \hline
       \textbf{Model} & \makecell{\textbf{Average} \\ \textbf{SE regret}} & \makecell{\textbf{Average} \\ \textbf{Spectral} \\ \textbf{Efficiency}} & \makecell{\textbf{\% of time} \\ \textbf{steps with} \\ \textbf{optimal action}} \\ \hline
       Exhaustive Search & 0 & 7.154 & 100\% \\ \hline
       Random action selection & 1.7751 & 5.378 & 1.90\% \\ \hline
       DRL - 1 epoch & 0.3424 & 6.811 & 40.90\% \\ \hline
       TRL - 1 epoch & 1.012 & 6.142 & 12.90\% \\ \hline
       DRL - 100 epoch (1) & 0.0854 & 7.068 & 33.10\% \\ \hline
       TRL - 100 epoch (2) & \textbf{0.0430} & \textbf{7.111} & \textbf{58.30\%} \\ \hline
       
   \end{tabular}}
   \label{Table:scenario44:summary}
   \vspace{-2ex}
\end{table}

The single UE scenario allows us to isolate and evaluate the performance of the multi-modal transformer-enhanced contextual bandit algorithm without the added complexity of multi-user interference.
Table \ref{Table:scenario44:summary} compares the performance of the models for Scenario 44 test data, where a single UE is mobile. 
In a single epoch, the DRL approach outperforms the TRL method. This is because both transformers and RL algorithms necessitate substantial data for effective training. However, when the number of epochs is increased to 100, the TRL achieves an average SE regret of 0.0430, as compared to 0.0854 achieved by DRL. This indicates that, with adequate data, the TRL is 49.6\% closer to the optimal value compared to the DRL.

\begin{figure}
    \centering
    \includegraphics[width=0.8\linewidth, height = 3.7cm, trim=44pt 16pt 49pt 40pt, clip]{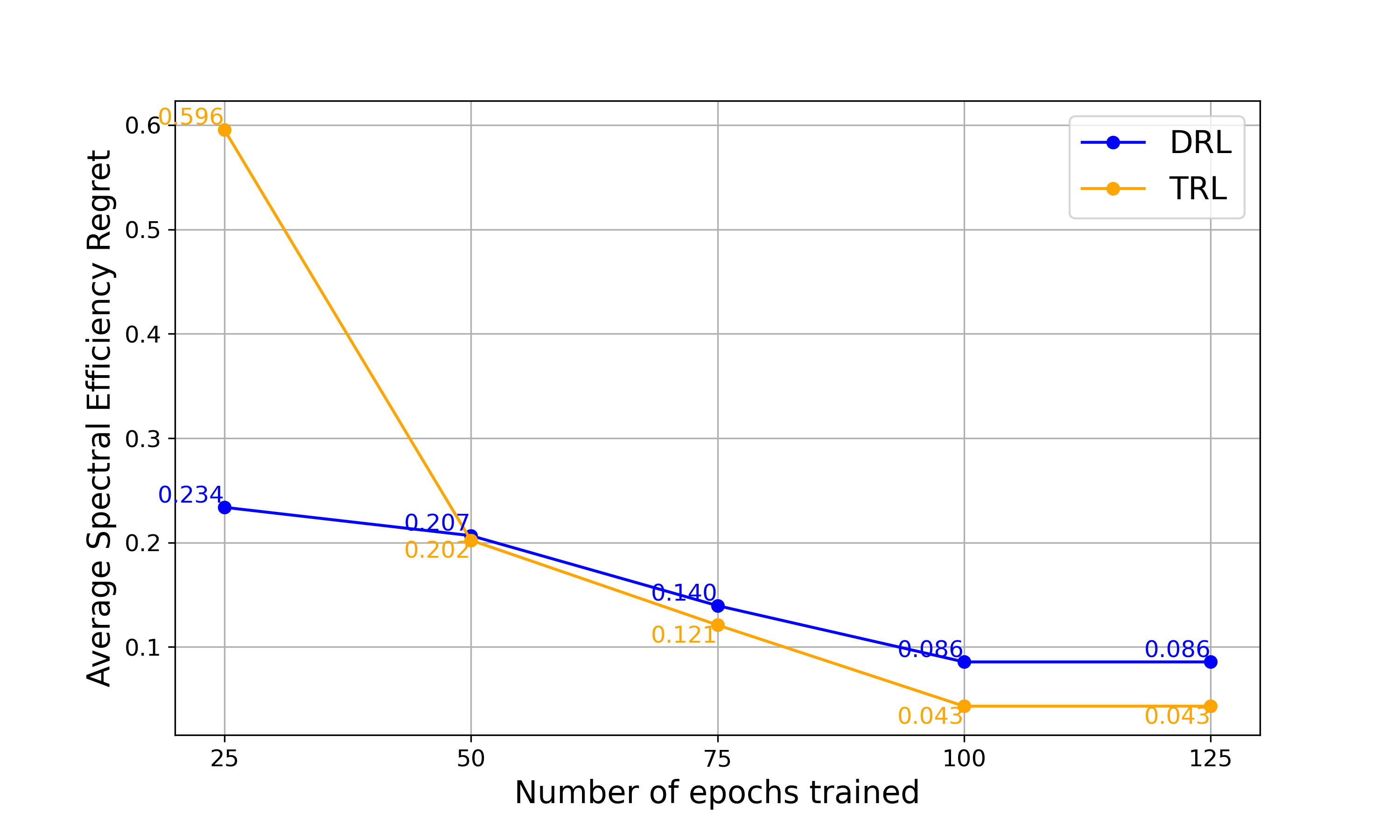}
    \caption{Average SE regret as a function of epochs.}
    \label{fig:Scenario44:Epochs}
\end{figure}

Fig. \ref{fig:Scenario44:Epochs} illustrates the performance of DRL and TRL as the models are trained for different numbers of epochs. It is evident that for 50 or more epochs, TRL surpasses DRL, achieving lower regret. With sufficient training, TRL approaches the optimal policy more closely than DRL.

\begin{figure}
     \centering
     \begin{subfigure}[b]{0.24\textwidth} 
         \centering
    \includegraphics[width=\textwidth,height=3.75cm,trim=20pt 5pt 42pt 33pt, clip]{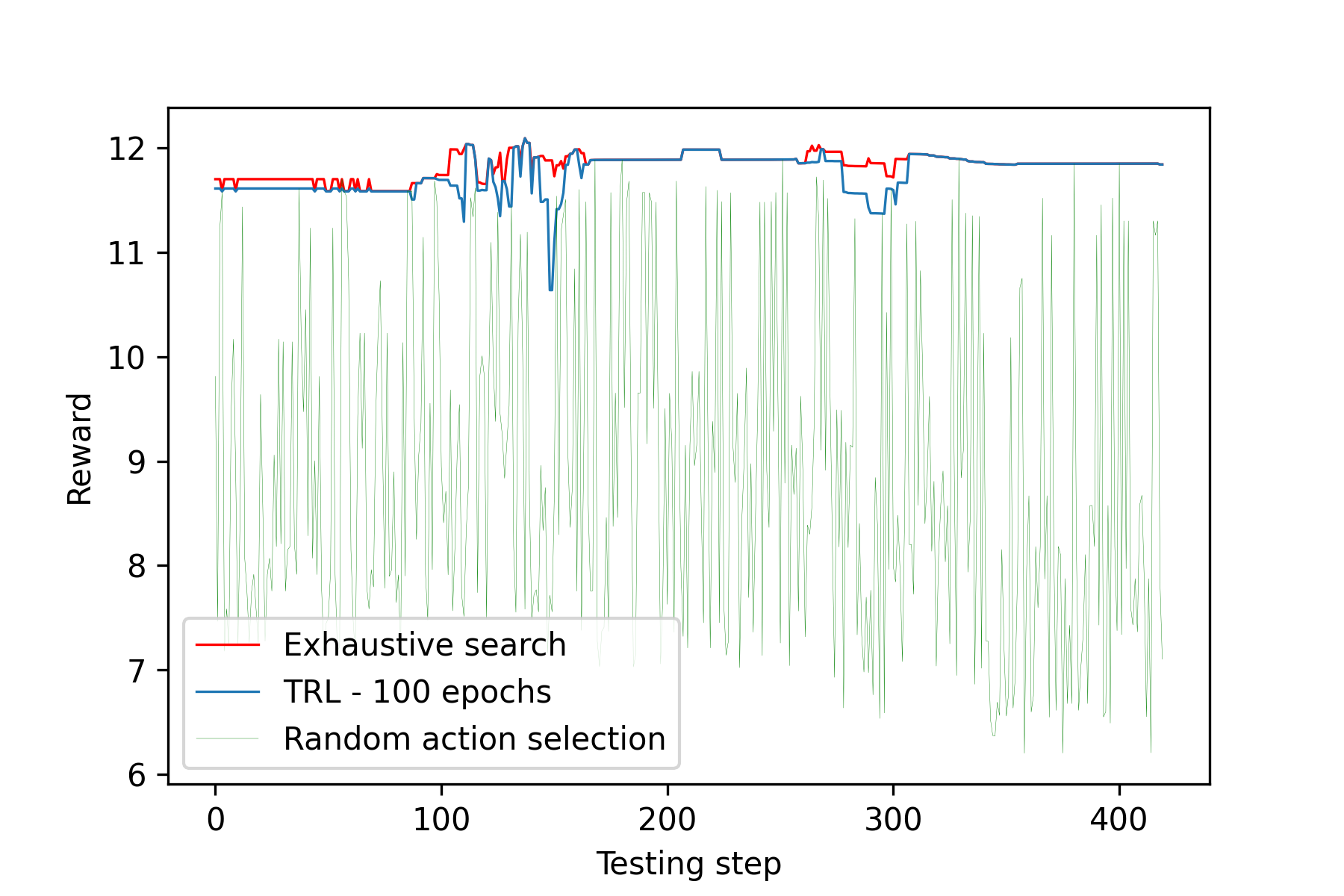}
         \caption{Reward}
         \label{fig:scenario44:reward}
     \end{subfigure}
     \begin{subfigure}[b]{0.24\textwidth}
         \centering
         \includegraphics[width=\textwidth,height=3.75cm,trim=20pt 5pt 42pt 33pt, clip]{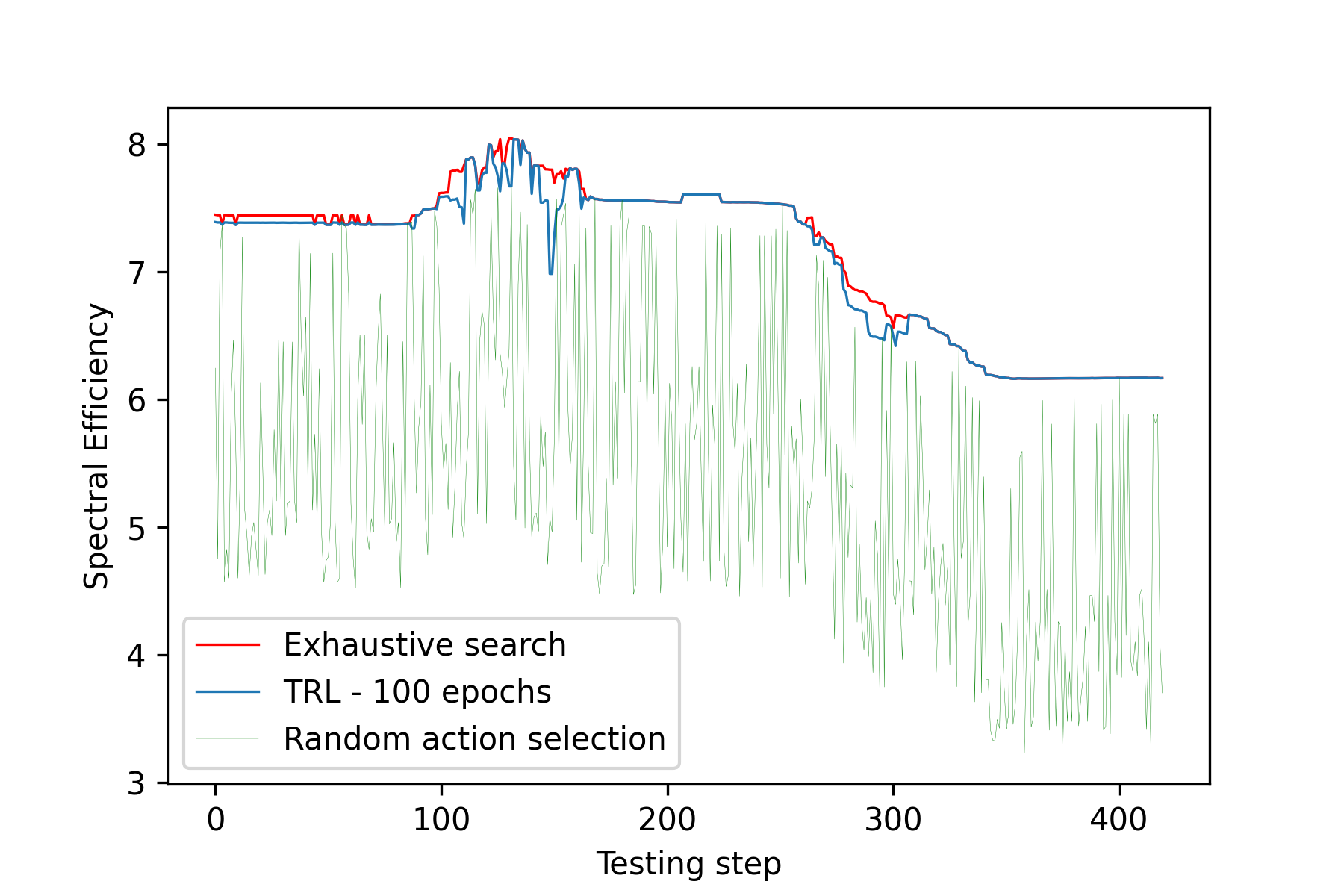}
         \caption{Spectral Efficiency}
         \label{fig:scenario44:se}
     \end{subfigure}
        \caption{Reward and Spectral Efficiency for each time step during the testing stage.}
        \label{fig:scenario44:results}
\end{figure}

Fig. \ref{fig:scenario44:reward}, and Fig. \ref{fig:scenario44:se} show the reward and spectral efficiency for each time step during the testing stage for the TRL model trained for 100 epochs. 
In Fig. \ref{fig:scenario44:se}, the optimal spectral efficiency varies as the UE moves around in the environment.
However, the reward curve (Fig. \ref{fig:scenario44:reward}) is smoother than the SE curve (Fig. \ref{fig:scenario44:se}). As explained in Section \ref{Section:ContextualDecisionFramework}, adding a distance-based factor to the reward design smooths the reward function and reduces agent bias for specific actions, leading to stable convergence.
It can be seen that the TRL model performs very close to the exhaustive search algorithm, showing it is effective in finding near-optimal solutions. 

\vspace{-1ex}
\subsection{Scenario 42 (Multiple UE) results}

\begin{table} [h]
\centering
\begin{tabular}{|c|c|c|}
\hline
\textbf{Model} & \makecell{\textbf{Average of sum}  \\ \textbf{Spectral Efficiency}} & \makecell{\textbf{Average} \\ \textbf{SE regret}} \\ \hline
Exhaustive Search & 5.427 & 0 \\ \hline
Random action selection & 2.131 & 3.2968 \\ \hline
DRL - 1 epoch & 3.78 & 1.6475 \\ \hline
DRL - 100 epoch & 4.655 & 0.772 \\ \hline
TRL - 100 epoch & \textbf{4.728} & \textbf{0.6995} \\ \hline
\makecell{TRL - 100 epochs \\ All scenarios used for training} & 4.561 & 0.8666 \\ \hline
\makecell{TRL - 100 epochs \\ Single agent for both users} & 4.525 & 0.902 \\ \hline
\end{tabular}
\caption{Average SE regret and sum spectral efficiency for different models in Scenario 42 test data.}
\label{Table:scenario42:first}
\end{table}

Next, we proceed to Scenario 42, in which two UEs traverse the environment. 
Compared with the single agent scenario, the average SE regret is higher because the users need to collaboratively choose actions to minimize mutual interference.

Table \ref{Table:scenario42:first} indicates that after training for 100 epochs, the TRL method achieves a 9.4\% improvement in proximity to the exhaustive search benchmark compared to the DRL method. Additionally, when all scenarios were utilized for training over 100 epochs, the resulting regret was 0.8666. This suggests that a more targeted training approach might be necessary to achieve lower regret values. Furthermore, employing a single agent for both users yielded an average SE regret of 0.902. The single-agent approach struggles to adequately address the distinct needs or interactions of multiple users, resulting in suboptimal performance compared to more specialized or individualized strategies. 

\subsection{Transfer Learning from single UE to multiple UE}

Finally, we show the results when we transfer the models from single agent Scenario 44 to multiple agent Scenario 42. We take the DRL model and TRL models trained for 100 epochs on Scenario 44 (marked as (1) and (2), respectively, in Table \ref{Table:scenario44:summary}). First, we directly transfer the model to the two agents in Scenario 42 and test the results. Next, we train the model for 1 epoch to measure further improvement.

\begin{table}[h]
    \centering
    \begin{tabular}{|c|c|c|}
        \hline
        & \textbf{TL - DRL} & \textbf{TL - TRL} \\
        \hline
        Use Scenario 44 model directly & 0.9174 & 0.7621 \\
        \hline
        Train for 1 epoch & 0.8714 & \textbf{0.5618} \\
        \hline
    \end{tabular}
    \caption{Average SE regret with transfer learning.}
    \vspace{-2ex}
\end{table}

For DRL, applying the model trained on Scenario 44 directly to the agents in Scenario 42 results in a regret value of 0.9174. However, conducting a single epoch of training on this scenario reduces the regret to 0.8714. Meanwhile, for TRL, utilizing the model from Scenario 44 without additional training yields an average SE regret of 0.7621. After one epoch of training, this regret decreases to 0.5618, representing the least regret achieved. This demonstrates that a transformer-based model can effectively learn a more generalized representation, enabling it to perform well across different scenarios. Notably, even a single training epoch can substantially enhance its performance.

\begin{figure}
    \centering
    \includegraphics[width=0.85\linewidth, trim=22pt 5pt 25pt 34pt, clip]{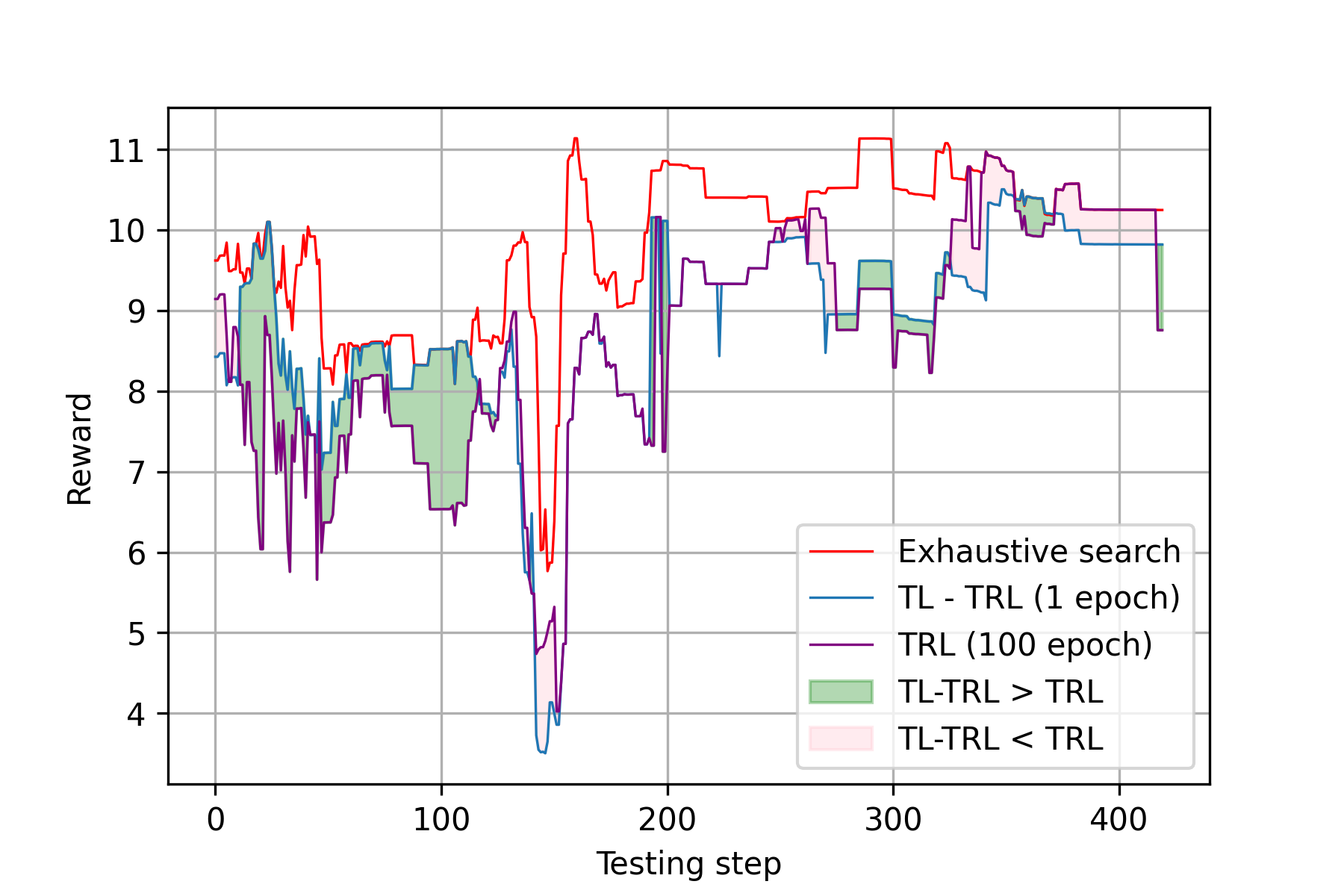}
    \caption{Reward comparison for each test time step between TRL model (100 epoch) and TL-TRL model (1 epoch).
    }
    \label{fig:scenario42_tl_reward}
\end{figure}
Fig. \ref{fig:scenario42_tl_reward} compares the test reward between the TRL model trained for 100 epochs (purple) and the TL-based TRL model for 1 epoch (blue). It can be seen that the TL-based model is closer to the exhaustive search model, indicating superior performance. The average sum spectral efficiency of the TL-TRL model after one epoch is 4.865, improving from 4.728 of the TRL model with 100 epochs.

\section{Conclusion} \label{Section:Conclusion}
\vspace{-1ex}
This work introduces an innovative approach to beam selection in ISAC systems by integrating MMT with a multi-agent contextual bandit algorithm and TL. The framework effectively utilizes ISAC sensing data to enhance communication performance, achieving high spectral efficiency in complex indoor environments. Our experimental results, validated on the DeepSense 6G dataset, show that the proposed model not only enhances beam prediction accuracy but also generalizes well across various scenarios, highlighting its potential for dynamic 6G networks. Furthermore, utilizing TL, we were able to reduce training time and improve the regret of the TRL model. 
In the future, we plan to explore outdoor scenarios.

\vspace{-1ex}
\balance
\bibliographystyle{IEEEtran} 
\bibliography{icc_paper}
\balance \balance


\end{document}